\RequirePackage{fix-cm}
\documentclass{svjour3}                     
\smartqed  
\usepackage{amsmath}
\usepackage{amssymb}
\usepackage{graphicx}
\usepackage{diagbox}
\usepackage{makecell}

\begin{document}

\title{A translation invariant bipolaron in the Holstein model and superconductivity.}

\author{Victor Lakhno}


\institute{Victor Lakhno \at
              Institute of Mathematical Problems of Biology, Keldysh Institute of Applied
Mathematics, Russian Academy of Sciences, Pushchino, Moscow Region, 142290, Russia. \\
              \email{lak@impb.psn.ru}
}

\date{ }

\maketitle

\begin{abstract}
Large-radius translation invariant (TI) bipolarons are considered in a one-dimensional
Holstein molecular chain. Criteria of their stability are obtained. The energy of
a translation invariant bipolaron is shown to be lower than that of a bipolaron
with broken symmetry. The results obtained are applied to the problem of superconductivity in 1D-systems.
It is shown that TI-bipolaron mechanism of Bose-Einstein condensation
can support superconductivity even for infinite chain.

\keywords{Delocalized \and broken symmetry \and strong coupling \and canonical transformation \and Hubbard Hamiltonian\and Bose condensate}
\end{abstract}

\section {Introduction}

The problem of possible existence of superconductivity in low-dimensional
molecular systems has long been of interest to researchers \cite{lit5}-\cite{lit10}. Presently,
it is believed that this phenomenon may occur via a bipolaron mechanism.
In three-dimensional systems a bipolaron gas is thought to form a Bose
condensate possessing superconducting properties. It is well known that in
one-and two-dimensional systems the conditions for bipolarons formation are
more favorable than in three-dimensional ones. The main problem in this
regard is the fact that in one- and two-dimensional systems Bose-condensation
is impossible \cite{lit-7-Ginzburg}.

In papers \cite{lit32}-\cite{Lak-2013} a concept of translation invariant polarons and bipolarons
was introduced. Under certain conditions these quasiparticles can possess
superconducting properties even if they do not form a Bose condensate.
Papers \cite{lit32}-\cite{Lak-2013} dealt with three-dimensional translation-invariant polarons
and bipolarons. In the context of the aforesaid it would be interesting to
consider the conditions under which translation invariant bipolarons arise in
low-dimensional systems. Here the results of \cite{lit32}-\cite{Lak-2013} are applied to the quasione-
dimensional case corresponding to the Holstein model of a large-radius
polaron.

In recent years increased interest in physics of 1D polarons and 1D bipolarons has been considerably provoked by the development of a lot of new materials, such as metal-oxyde ceramics with layered ($La_2$ ($Sr,Br$) $CuO_4$ and $(Bi,Tl)_2$ $(Sr,Ba)_2$ $CaCuO_8$) or layered-chain ($Y$ $Ba_2Cu_3O_7$) structure, demonstrating high-temperature superconductivity \cite{lit1}-\cite{lit4}, chain organic (polyacetylene) and inorganic ($(SN)_x$) polymers, quasi-one-dimensional conducting compounds where charge transfer takes place (TTF – TCNQ), etc. \cite{lit5}-\cite{lit10kl}. Much the same as 1D systems can be materials with huge anisotropy where polarons or bipolarons can emerge \cite{lit10s}-\cite{lit10kl}. Development of DNA-based nanobioelectronics \cite{lit11}, \cite{lit12} is also closely related with calculation of polaron and bipolaron properties in one-dimensional molecular chains \cite{lit13}-\cite{lit16}. Despite great theoretical efforts, many problems of polaron physics have not been solved yet.

One of the central problems of polaron physics is that of spontaneous breaking of symmetry of the "electron + lattice" system. In most papers on polaron physics, following initial Landau hypothesis \cite{lit-Landau} valid for classical lattice,  (see books and reviews \cite{lit17}-\cite{lit25}) it is thought that at rather a large
 coupling an electron deforms a lattice so heavily that it becomes self-trapped in the deformed region. In this case the initial symmetry of the Hamiltonian is broken: an electron passes on from the delocalized state having the Hamiltonian symmetry to the localized "self-trapped" state with broken symmetry. This problem is still more actual for bipolarons since a bipolaron state can arise only in the case of large values of the coupling constant.

As showed in Ref.~\cite{lit26} for 1D Holstein polaron in a continuum limit for all the values of the coupling constant, the minimum of its energy in quantum lattice is reached in the class of delocalized wave functions. So in Ref.~\cite{lit26} it is shown that in the case of a strong-coupling polaron, symmetry is not broken and a self-trapped state is not formed.

In this paper the results of paper \cite{lit26} are generalized to the case of 1D bipolaron.

In \S2 we present known exact results for a polaron and bipolaron with broken translation invariance in the Holstein continuum model in the strong coupling limit when Coulomb interaction between electrons is lacking. In the general case, when the Coulomb interaction takes place, the properties of the bipolaron ground state are illustrated with the use of a variational approach in which the localized wave function of the exact solution without Coulomb interaction is used as a probe one. The results obtained are used to present the criteria of the bipolaron stability.

In \S3 a translation invariant bipolaron theory is constructed. The wave function of such a bipolaron is delocalized. In the strong coupling limit the functional of the bipolaron total energy is derived.

In \S4 to study the minimum of the total energy a direct variational method is used. It is shown that, as distinct from a bipolaron with broken symmetry, a translation invariant bipolaron exists for all the values of the Coulomb repulsion constant. The regions where a translation invariant bipolaron is stable relative to its decay into two individual polarons are found. It is shown that for all the values of the Coulomb repulsion parameter, the energy of a translation invariant bipolaron is lower than that of a bipolaron with spontaneously broken symmetry.

In \S5 we analyze solutions of the equations for the translation invariant bipolaron (below TI-bipolaron) spectrum. It is shown that the spectrum has a gap separating the ground state of a TI-bipolaron from its excited states which form a quasicontinuous spectrum. The concept of an ideal gas of TI-bipolarons is substantiated.

With the use of the spectrum obtained, in \S6 we consider thermodynamic characteristics of an ideal gas of TI-bipolarons. For various values of the parameters, namely phonon frequencies, we calculate the values of critical temperatures of Bose condensation, latent heat of transition into the condensed state, heat capacity and heat capacity jumps at the point of transition.

In \S7 we compare the results for continuum and discrete models.

In \S8 we discuss the results obtained.

\section {Bipolarons with broken translation invariance in the Holstein model in the strong coupling limit.}

According to Ref.~\cite{lit26}--\cite{lit28} Holstein Hamiltonian in a one-dimensional chain in a continuum limit has the form:

\begin{eqnarray}\label{1}
H = -\frac{1}{2m}\Delta _{x_{1}}-\frac{1}{2m}\Delta _{x_{2}}+
\sum_{k}\left[V_{k}\left(e^{ikx_1}+e^{ikx_2}\right)a_k+h.c.\right]+ \\ \nonumber
\sum_k\hbar\omega^0_ka^+_ka_k+U\left(x_1-x_2\right),\
V_k=\frac{g}{\sqrt{N}},\ \omega ^0_k=\omega _0,
\end{eqnarray}
where $a^+_k$, $a_k$ are operators of the phonon field, $m$ is the electron effective mass,
$\omega^0_k$ is the frequency of optical phonons, $g$ is the constant of electron-phonon interaction,
$N$ is the number of atoms in the chain, $U(x_1-x_2)$ is the Coulomb repulsion between  electrons
depending on the difference of electron coordinates which will be taken to be:

\begin{eqnarray}\label{2}
U\left(x_1-x_2\right)= \Gamma \delta \left(x_1-x_2\right)
\end{eqnarray}
where $\Gamma$ is a certain constant, $\delta (x)$ is a delta function. In the case of broken translation
invariance the bipolaron state is described by localized wave functions $\Psi = \Psi (x_1,x_2)$
and in the strong coupling limit the functional of the total energy $\bar{H}=\left\langle \Psi\left|H\right|\Psi\right\rangle$ is written as \cite{lit29}:

\begin{eqnarray}\label{3}
\bar{H}=-\frac{1}{2m}\sum_{i=1,2}\left\langle \Psi\left|\Delta _{x_i}\right|\Psi\right\rangle-
\sum\frac{V^2_k}{\hbar \omega_0}\left\langle \Psi\left|e^{ikx_1}+e^{ikx_2}\right|\Psi\right\rangle ^2+ \\ \nonumber
\left\langle \Psi\left|U\left( x_1-x_2\right)\right|\Psi\right\rangle
\end{eqnarray}
The exact solution of problem (3) is a complicated computational problem \cite{lit30a}-\cite{lit30e}.
For the purposes of this section,
however, it will suffice to illustrate the properties of the ground state of a bipolaron with broken
symmetry with the use of a direct variational method. Towards this end let us choose the probe function
$\Psi = \Psi (x_1,x_2)$ in the form $\Psi (x_1,x_2)=\varphi(x_1)\varphi(x_2)$. Notice that this choice of
the probe function corresponds to the exact solution of problem (3) for $U=0$, i.e. in the absence of the
Coulomb interaction between electrons.

As a result, from (3) we get the functional of the ground state energy:

\begin{eqnarray}\label{4}
\bar{H}=\frac{1}{m}\int\left|\nabla _x\varphi (x)\right|^2dx-\left(\frac{4g^2a_0}{\hbar \omega _0}-\Gamma\right)\int\left|\varphi (x)\right|^4dx,
\end{eqnarray}
where $a_0$ is the lattice constant.
Variation of (4) with respect to  $\varphi(x)$, the normalization requirement being met, leads to Schroedinger equation:

\begin{eqnarray}\label{5}
\frac{\hbar^2}{m}\Delta _x\varphi + 2\left(\frac{4g^2a_0}{\hbar \omega _0}-\Gamma\right)\left|\varphi\right|^2\varphi + W\varphi=0,
\end{eqnarray}
whose solution has the form:
\begin{eqnarray}\label{6}
\varphi (x)=\pm\left( \sqrt{2r}ch\frac{x-x_0}{r}\right)^{-1},\ r=\frac{2\hbar ^2}{m}\frac{1}{\left({(4g^2a_0)}/{(\hbar \omega _0)}-\Gamma\right)},\\ \nonumber
W=-\frac{1}{2}\left(\frac{4g^2a_0}{\hbar \omega _0}-\Gamma\right)^2\frac{m}{2\hbar ^2},\
E_{bp}=-\frac{1}{6}\left(\frac{4g^2a_0}{\hbar \omega _0}-\Gamma\right)^2
\frac{m}{2\hbar ^2},
\end{eqnarray}
where $x_0$ is an arbitrary constant, $E_{bp}=min\;\bar{H}$ is the energy of the bipolaron ground state.
Notice, that the polaron state energy $E_p$ in the case under consideration is \cite{lit26}:

\begin{eqnarray}\label{7}
E_P=-\frac{1}{6}\left(\frac{g^2a_0}{\hbar\omega _0}\right)^2\frac{m}{\hbar ^2}.
\end{eqnarray}
Let us introduce the notation:
\begin{eqnarray}\label{8}
\gamma=\Gamma\hbar\omega _0/a_0g^2.
\end{eqnarray}
From (6) it follows that for:
\begin{eqnarray}\label{9}
\gamma>4
\end{eqnarray}
the existence of the bipolaron state is impossible. In the case of:
\begin{eqnarray}\label{10}
2<\gamma<4
\end{eqnarray}
the metastable bipolaron state will decay into individual polaron states. As:
\begin{eqnarray}\label{11}
\gamma<2
\end{eqnarray}
the bipolaron state will be stable. Notice that the choice of more complex probe functions \cite{lit30a} has no effect on the qualitative picture presented, changing only the numerical coefficients in relations (9) - (11).

In view of an arbitrary position of the bipolaron center of mass  $x_0$, the bipolaron state discussed has an infinite degeneracy and can move along the chain. Any arbitrarily small violation of the chain leads to elimination of the degeneration and localization of the bipolaron state on defects with attracting potential.  A qualitatively different situation arises in the case of a translation invariant bipolaron considered below.

\section { Translation invariant bipolaron theory.}

To construct a translation invariant bipolaron theory in the Holstein model, in Hamiltonian (1) we pass on to coordinates of the center of mass. In this system Hamiltonian (1) takes the form:

\begin{eqnarray}\label{12}
H=-\frac{\hbar^2}{2M}\Delta_{R}-\frac{\hbar^2}{2\mu}\Delta_{r}+
\sum_k 2V_k cos\frac{kr}{2}\left(e^{ikR}a_k+h.c.\right)+ \\ \nonumber
\sum_k\hbar\omega ^0_k a^+_k a_k+U(r),\\ \nonumber
R=\left(x_1+x_2\right)/2,\ r=x_1-x_2,\ M=2m,\ \mu=m/2.
\end{eqnarray}
In what follows we will use units, putting  $\hbar=1$, $\omega _0=1$, $M=1$ (accordingly  $\mu=1/4$).

The coordinate of the center of mass $R$ in Hamiltonian (2) can be eliminated via Heisenberg canonical transformation \cite{lit30}:

\begin{eqnarray}\label{13}
\hat{S}_1=exp\left\{-i\sum_kka^+_ka_kR\right\} .
\end{eqnarray}
As a result, the transformed Hamiltonian: $\tilde{H}=\hat{S}^{-1}_1H\hat{S}_1$ is written as:
\begin{eqnarray}\label{14}
\tilde{H}=-2\Delta_r+\sum_k2V_k cos \frac{kr}{2}(a^+_k+a_k)+
\sum_ka^+_ka_k+U(r)+\\\nonumber
\frac{1}{2}\left(\sum_ka^+_ka_k\right)^2
\end{eqnarray}
From (14) it follows that the exact solution of the bipolaron problem is determined by the wave function
$\Psi(r)$ which depends only on the relative coordinates $r$ and, therefore, is automatically translation
invariant. It corresponds to the state delocalized over the coordinates of the center of mass of two electrons.

Averaging Hamiltonian (14) over  $\Psi(r)$, we will write the averaged Hamiltonian  as: $\bar{\tilde{H}}$ (15)

\begin{eqnarray}\label{15}
\bar{\tilde{H}} = \bar{T}+\sum_k \bar{V}_k(a^+_k+a_k)+
\sum_ka^+_ka_k+\frac{1}{2}\left(\sum_ka^+_ka_k\right)^2+\bar{U},\\\nonumber
\bar{V}_k=2V_k\left\langle \Psi\left| cos \frac{kr}{2}\right|\Psi\right\rangle,
\bar{U}=\left\langle \Psi\left|U(r)\right|\Psi\right\rangle,
\bar{T}=-2\left\langle \Psi\left|\Delta _r\right|\Psi\right\rangle
\end{eqnarray}
Subjecting Hamiltonian (15) to Lee-Low-Pines transformation \cite{lit31}:
\begin{eqnarray}\label{16}
\hat{S}_2=exp\left\{\sum_k f_k(a_k-a^+_k)\right\},
\end{eqnarray}
we get:
\begin{eqnarray}\label{17}
\tilde{\tilde{H}}=\hat{S}^{-1}_2\bar{\tilde{H}}\hat{S}_2\ \ \ \ \tilde{\tilde{H}}=H_0+H_1
\end{eqnarray}
where:
\begin{eqnarray}\label{18}
H_0=\bar{T}+2\sum_k\bar{V}_kf_k+\sum_kf^2_k+\frac{1}{2}\left(\sum_kkf_k\right)^2+\bar{U}+{\cal{H}}_0
\end{eqnarray}

\begin{eqnarray}\label{19}
{\cal {H}}_0=\sum_k\omega_ka^+_ka_k+
\frac{1}{2}\sum_{k,k'}kk'f_kf_{k'}\left(a_ka_{k'}+a^+_ka^+_{k'}+a^+_ka_{k'}+a^+_{k'}a_k\right),
\end{eqnarray}

\begin{eqnarray}\label{20}
H_1=\sum_k\left(V_k+f_k\omega_k\right)\left(a_k+a^+_k\right)+
\sum_{k,k'}kk'f_{k'}\left(a^+_ka_ka_{k'}+a^+_ka^+_{k'}a_k\right)+\\\nonumber
\frac{1}{2}\sum_{k,k'}kk'a^+_ka^+_{k'}a_ka_{k'},
\end{eqnarray}

\begin{eqnarray}\label{21}
\omega_k=\omega_0+\frac{k^2}{2}+k\sum_{k'}k'f^2_{k'}.
\end{eqnarray}
According to Ref.~\cite{lit32}, contribution of $H_1$ into the energy vanishes if the eigen function of Hamiltonian ${\cal {H}}_0$
transforming the quadratic form ${\cal{H}}_0$ to the diagonal one, is chosen properly. Diagonalisation of ${\cal{H}}_0$
 leads to the total energy of the addition $\Delta E$:
\begin{eqnarray}\label{22}
\Delta E=\frac{1}{2}\sum_{k}\left(\nu_k-\omega_k\right)=-\frac{1}{8\pi i}\int_{c}\frac{ds}{\sqrt{s}}ln D(s),
\end{eqnarray}
where  $\nu _k$ are phonon frequencies renormalized by the interaction with the electron. The contour of integration $c$
involved in (22) is the same as in Ref.~\cite{lit32},~\cite{lit26}. In the one-dimensional case under consideration:
\begin{eqnarray}\label{23}
D(s)=1-\frac{1}{\pi}\int^{\infty}_{-\infty}\frac{k^2f_k\omega_k}{s-\omega^2_k}dk
\end{eqnarray}
Repeating calculations carried out in Ref.~\cite{lit32},~\cite{lit26} in the strong coupling limit, we express $\Delta E$ as:
\begin{eqnarray}\label{24}
\Delta E=\frac{1}{4\pi}\int^{\infty}_{-\infty}\frac{k^2f^2_kdk}{2(1+Q)}+\\\nonumber
\frac{1}{4\pi^2}\iint ^{\infty}_{-\infty}\frac{k^2f^2_kp^2f^2_p\omega_p(\omega_k\omega_p+\omega_k(\omega_k+\omega_p))+1}
{(\omega_k+\omega_p)^2(\omega^2_p-1)\left|D_+(\omega^2_p)\right|^2}dpdk,\\\nonumber
D_+(\omega^2_p)=1+\frac{1}{\pi}\int^{\infty}_{-\infty}\frac{f^2_kk^2\omega_kdk}{\omega^2_k-\omega^2_p-i\epsilon},\\\nonumber
Q=\frac{1}{\pi}\int^{\infty}_{-\infty}\frac{k^2f^2_k\omega_kdk}{\omega^2_k-1}
\end{eqnarray}
Finally, with the use of (18) and (19) the bipolaron total energy $E_{bp}$ is written as:

\begin{eqnarray}\label{25}
E_{bp}=\Delta E+2\sum_k\bar{V}_kf_k+\sum_kf^2_k+\bar{T}+\bar{U}
\end{eqnarray}

\section { Variational calculation of the bipolaron state.}

We could have derived an exact equation for determining the bipolaron energy by varying (25) with respect to $\Psi$ and $f_k$.
 The quantities $\Psi$  and  $f_k$ obtained as solutions of this equation, being substituted into (25)
 determine the bipolaron total energy $E_{bp}$. Since finding a solution of the equation obtained by variation
 of $E_{bp}$ is rather a complicated procedure, we will use the variational approach. To this end let us choose
 the probe functions $\Psi$  and  $f_k$  in the form:

\begin{eqnarray}\label{26}
\Psi(r)=\left(\frac{2}{\pi}\right)^{1/4}\frac{1}{\sqrt{l}}e^{-r^2/l^2},
\end{eqnarray}

\begin{eqnarray}\label{27}
f_k=-Nge^{-k^2/2a^2},
\end{eqnarray}
where  $N$, $l$, $a$ are variational parameters. As a result, after minimization of (25) on $N$, the bipolaron energy will be:
\begin{eqnarray}\label{28}
E_{bp}=\frac{ma^{2}_{0}}{\hbar^2}\frac{g^4}{\hbar^2\omega^{2}_{0}}\;{min}_{(x,y)}E\left(x,y;\gamma\right),
\end{eqnarray}

\begin{eqnarray}\label{29}
E(x,y;\gamma)\approx 2\left(0.390625x^2+\frac{2}{y^2}-\frac{4x}{\sqrt{\pi}(1+x^2y^2/16)}+\sqrt{\frac{2}{\pi}}\frac{\gamma}{y}\right)
\end{eqnarray}

\begin{figure}
\includegraphics{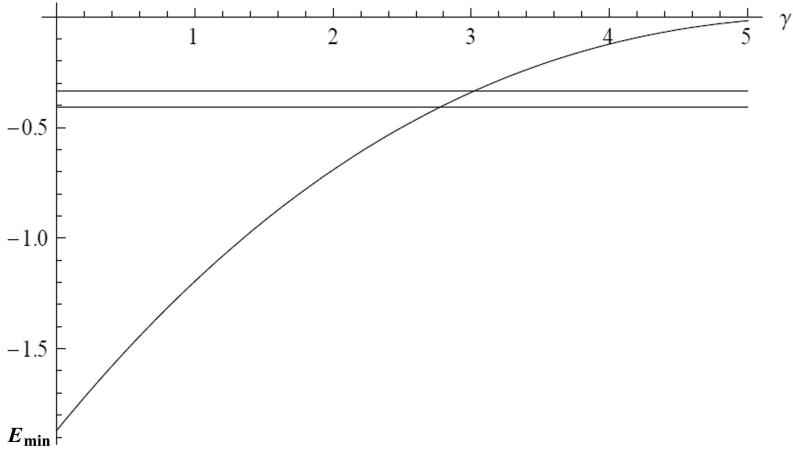}
\caption{The dependence of $E_{min}$ (29) on $\gamma$.}
\end{figure}
The expression for the bipolaron energy is given in dimension units. The results of minimization of function $E(x,y;\gamma)$ with respect to dimensionless parameters $x,y$ are presented in Fig.~1 for various values of the parameter $\gamma$. Fig.~1 suggests that as distinct from a bipolaron with broken symmetry (inequality (9)), a translation invariant bipolaron exists for all the values of the parameter $\gamma$. In the region:
\begin{eqnarray}\label{30}
\gamma > 3.02
\end{eqnarray}
a translation invariant bipolaron is unstable relative to its decay into both individual polarons with spontaneously broken symmetry, i. e. Holstein polarons with the energy $2E_p=-(1/3)ma^2_0g^4/\hbar^4\omega^2_0$ (upper horizontal line in Fig.~1 in energy units  $ma^2_0g^4/\hbar^4\omega_0$) and translation invariant polarons with the energy $2E_p=-0.4074ma^2_0g^4/\hbar^4\omega^2_0\,$ \cite{lit26}
(lower horizontal line in Fig.~1). For:
\begin{eqnarray}\label{31}
2.775 < \gamma <3.02
\end{eqnarray}
a translation invariant bipolaron becomes stable relative to its decay into individual Holstein polarons, but remains unstable relative to decomposition into individual translation invariant polarons. For:
\begin{eqnarray}\label{32}
\gamma < \gamma _c=2.775
\end{eqnarray}
a translation invariant bipolaron becomes stable relative to its decay into two individual polarons. Notice that for $\gamma=0$, the energy of a translation invariant bipolaron is equal to: $E_{bp}=-1.87104\,ma^2_0g^4/\hbar^4\omega^2_0$, i.e. lies much lower than the exact value of the energy of a bipolaron with broken symmetry, which, according to (6) is equal to $E_{bp}=-(4/3)ma^2_0g^4/\hbar^4\omega^2_0$. The energy of a translation invariant bipolaron also lies below the variational estimate of the energy of a bipolaron with spontaneously broken symmetry (6) for all the values of $\gamma$ \cite{lit30a}.

The dimensionless parameters $x,y$ involved in (29) are related to the variational parameters
$a$ and $l$ (26), (27) as: $a=(2ma^2_0g^2/\hbar^3\omega_0)x$, $l=(\hbar^3\omega_0/2ma^2_0g^2)y$.
The parameter $l$ determine the characteristic size of the electron pair, i.e. the correlation length
$L(\gamma)$, whose dependence on $\gamma$ is given by the expression:

\begin{eqnarray}\label{33}
L(\gamma)=\frac{\hbar^2}{2ma^{2}_{0}}\frac{\hbar\omega_0}{g^2}y_{min} (\gamma).
\end{eqnarray}

\begin{figure}
\includegraphics{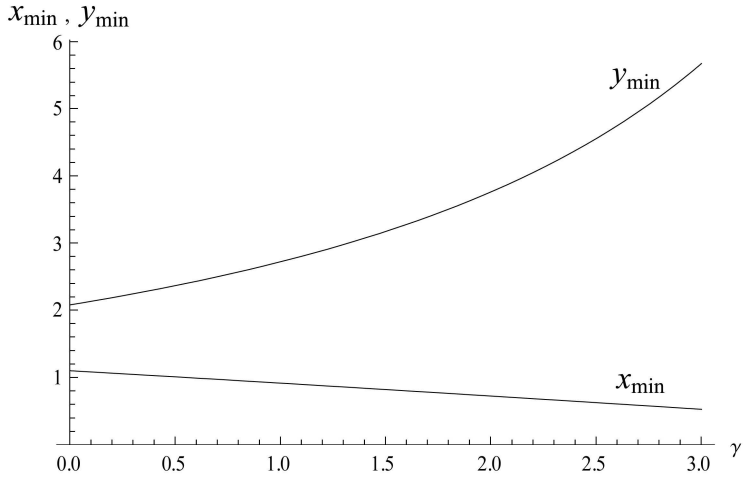}
\caption{The dependence of $x_{min}$, $y_{min}$  on $\gamma$.}
\end{figure}

The dependencies of $y_{min}$ and $x_{min}$ on $\gamma$ are presented in Fig.~2.

Fig.~2 suggests that the correlation length $L(\gamma)$ in the region of a bipolaron stability $0<\gamma<\gamma_c$
does not change greatly  and for its critical value $\gamma_c=2.775$ the quantity $L(\gamma)$ approximately three
times exceeds the value of $L(0)$, i.e. the correlation length in the absence of the Coulomb repulsion.
This qualitatively differs from the case of a bipolaron with broken symmetry for which the corresponding value,
according to (6), for $\gamma=\gamma_c$ turns to infinity.

\section {Spectrum of excited states.}

According to the results obtained in \cite{lit32}, \cite{lit35}, the spectrum of excited states of Hamiltonian (18), (19) is determined by the expression:
\begin{eqnarray}
\label {eq.1}
\tilde{\tilde{H}} = E_{bp}+\sum_k\nu_k\alpha^+_k\alpha_k
\end{eqnarray}

where $\alpha ^+_k$, $\alpha _k$ are operators in which quadric form  $H_0$ (19) is diagonal.
Operators  $\alpha ^+_k$, $\alpha _k$ can be considered as operators of birth and annihilation of TI-bipolarons in excited states obeying Bose commutation relations:
\begin{eqnarray}
\label {eq.2}
\left[\alpha_n,\alpha^+_{n'}\right] = \alpha_n\alpha^+_{n'} - \alpha^+_n\alpha_n =\delta_{n,n'}
\end{eqnarray}

Renormalized frequencies involved in (34), according to \cite{lit32}, \cite{lit35},  are determined by the equation for $s$:
\begin{eqnarray}
\label {eq.3}
1=2\sum_k\frac{k^2f_k\omega_k}{s-\omega^2_k}
\end{eqnarray}
solutions of which give the spectrum of  $s=\left\{\nu^2_k\right\}$ solutions.
	
It is convenient to present Hamiltonian (34) in the form:

\begin{eqnarray}
\label {eq.4}
\tilde{\tilde{H}} = \sum_{n=0,1,2}E_n\alpha^+_n\alpha_n
\end{eqnarray}

\begin{eqnarray}
\label {eq.5}
E_n=
\begin{cases}
E_{bp}, & n=0; \\
\nu_n=E_{bp}+\omega_0+\frac{k^2_n}{2}, &  n \neq 0.
\end{cases}
\end{eqnarray}

where $k_n$   for a discrete chain of atoms is equal to:
\begin{eqnarray*}
k_n=\pm\frac{2\pi(n-1)}{N_a},\ \ n=1,2...,N_a/2+1
\end{eqnarray*}
 $N_a$ is the number of atoms in the chain.

Let us prove the validity of (38).
The energy spectrum of TI-bipolarons, according to (36), reads:

\begin{eqnarray}
\label {eq.7}
F(s)=1
\end{eqnarray}

\begin{eqnarray}
\label {eq.8}
F(s)=2\sum_n\frac{k^2_nf^2_{k_n}\omega_{k_n}}{s-\omega^2_{k_n}}
\end{eqnarray}

It is convenient to solve equation (39) graphically (Fig.3)

\begin{figure}
\includegraphics{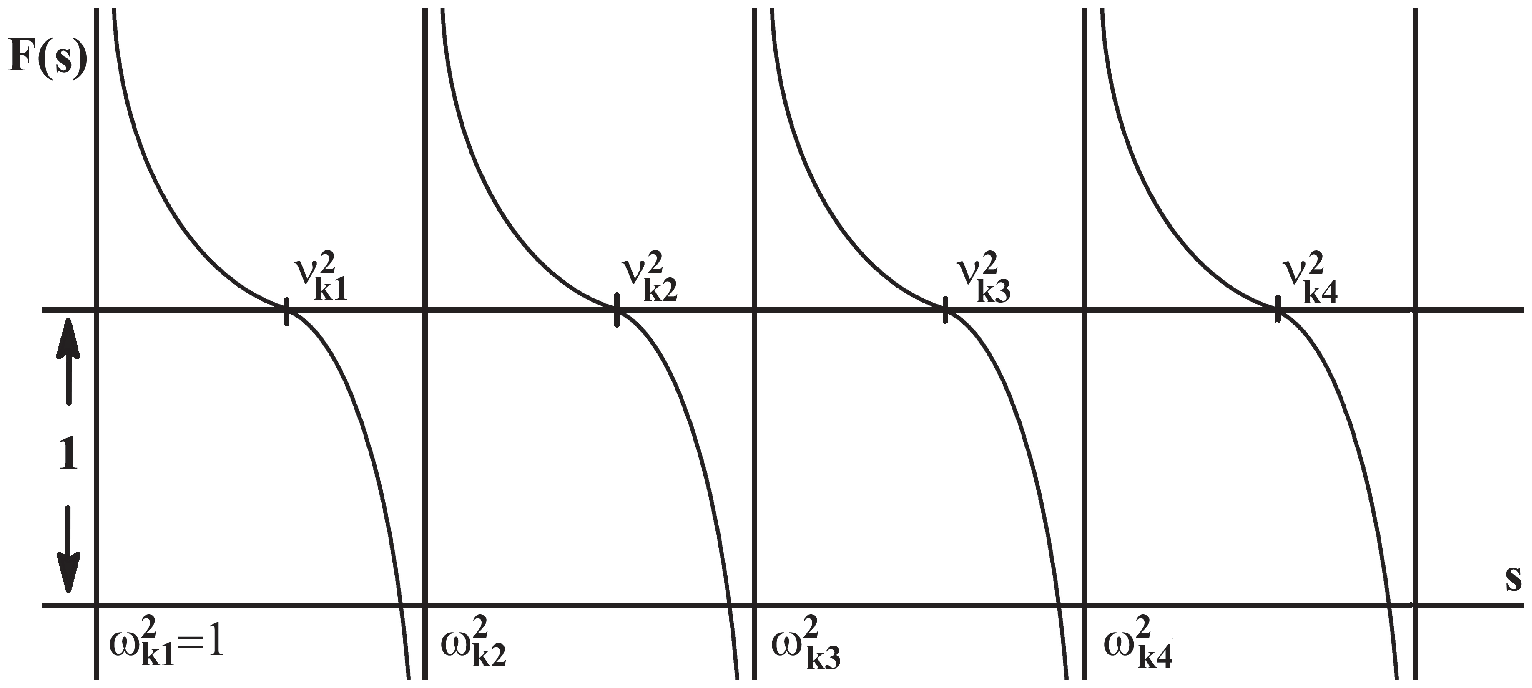}
\caption{Graphical solution of Eqs. (39), (40)}
\end{figure}

Fig.3 suggests that frequencies $\nu _{k_n}$ occur between the frequencies $\omega_{k_n}$   and  $\omega_{k_{n+1}}$.
Hence, the spectrum of $\nu _{k_n}$   as well as the spectrum of $\omega_{k_n}$   is quasi continuous in the continuum limit:  $\nu _{k_n}-\omega_{k_n}=0(N^{-1}_a)$, which proves the validity of (37), (38).

Therefore the spectrum of a TI-bipolaron has a gap between the ground state of $E_{bp}$   and the quasi continuum spectrum, which is equal to  $\omega _0$.

Below we will consider the case of low concentration of TI-bipolarons in the chain.  In this case they can be adequately considered as Bose-gas, whose properties are determined by Hamiltonian (37).

\section {Statistical thermodynamics of 1D gas of TI-bipolarons.}

Let us consider the rare (the pair correlation length is much smaller then the average distance between pairs) one-dimensional ideal Bose-gas of TI-bipolarons which is a system of  $N$  particles, occurring in a one-dimensional chain of length $L$. Let us write $N_0$ for the number of particles in the lower one-particle state, and  $N$ for the number of particles in higher states. Then:
\begin{eqnarray}
\label {eq.9}
N=\sum_{n=0,1,2,...}\bar{m}_n=\sum_n\frac{1}{e^{(E_n-\mu)/T}-1}
\end{eqnarray}

\begin{eqnarray}
\label {eq.10}
N=N_0+N',\ N_0=\frac{1}{e^{(E_0-\mu)/T}-1},\ \ N'=\sum_{n \neq 0}\frac{1}{e^{(E_n-\mu)/T}-1}
\end{eqnarray}

In expression for  $N'$ (42) we will replace summation by integration over quasi continuous spectrum (37), (38) and take $\mu=E_{bp}$.  As a result we will get from (41), (42) an expression for the temperature of Bose condensation  $T_c$:
\begin{eqnarray}
\label {eq.11}
C_{1D}=\Phi_{\tilde{\omega}}\left(T_c\right)
\end{eqnarray}

$$\Phi_{\tilde{\omega}}=\tilde{T}^{1/2}_{c}F_{1/2}\left(\frac{\tilde{\omega}}{\tilde{T}_c}\right),\
F_{1/2}(\alpha)=\int_0^{\infty}\frac{dx}{\sqrt{x}\left(e^{x+\alpha}-1\right)},$$

$$C_{1D}=2\sqrt{2}\pi\frac{n\hbar}{M^{1/2}\omega^{*1/2}},\
\omega^*=\frac{\omega_0}{\omega}, \tilde{T}_c=\frac{T}{\omega^*},$$

where  $n=N/L$. Fig.4 shows a graphical solution of equation (43) for the parameter values:  $M_e=2m=2m_0$, where  $m_0$ is the mass of a free electron in vacuum, $\omega ^*=5$ meV, ($\approx 58K$ ), $n=10^7$ cm$^{-1}$ and the values:  $\tilde{\omega}_1=0.2$; $\tilde{\omega}_2=1$;  $\tilde{\omega}_3=2$;  $\tilde{\omega}_4=10$;  $\tilde{\omega}_5=15$;    $\tilde{\omega}_6=20$
($C_{1D}=34.69$).

\begin{figure}
\includegraphics{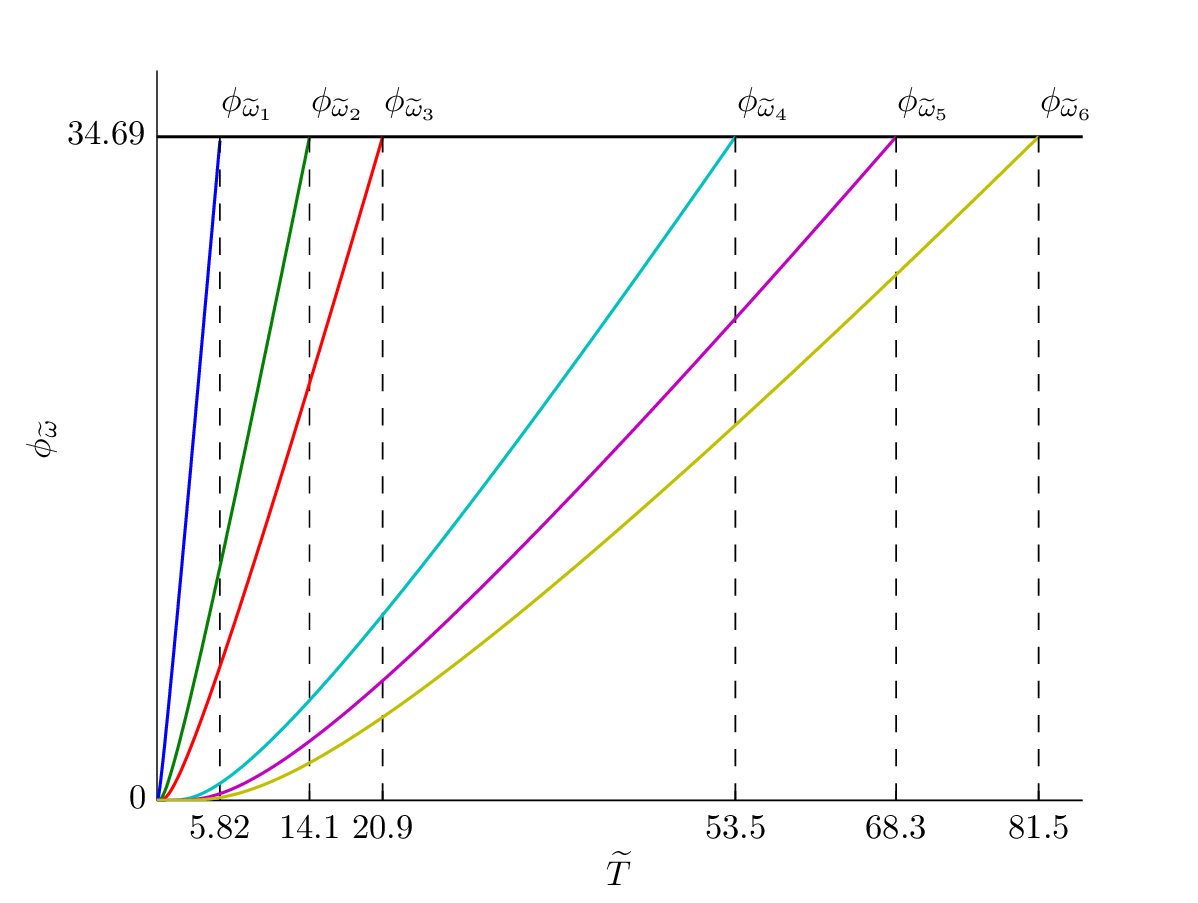}
\caption{Solutions of equation (43) $C_{1D}=34.69$  and $\tilde{\omega}_i=\left\{0.2;1;2;10;15;20\right\}$,
which correspond to   $\tilde{T}_{c_i}=\tilde{T}_{c_1}=5.823$;  $\tilde{T}_{c_2}=14.1$;  $\tilde{T}_{c_3}=20.87$;
$\tilde{T}_{c_4}=53.47$;  $\tilde{T}_{c_5}=68.33$; $\tilde{T}_{c_6}=81.5$.}
\end{figure}

Fig.~4 suggests that the critical temperature grows as the phonon frequency increases and is equal to zero for  $\omega =0$. The equality  $T_c=0$  for  $\omega =0$ corresponds to the known result, that Bose-condensation is impossible in ideal gas in a one-dimensional case.

Fig.~4 also suggests that it is just the increase in the concentration of TI-bipolarons which will lead to an increase in the critical temperature, while the increase in the electron mass  $m$ – to its decrease.

It follows from (41), (42) that:

\begin{eqnarray}
\label {eq.12}
\frac{N'(\tilde{\omega})}{N}=\frac{\tilde{T}^{1/2}}{C_{1D}}F_{1/2}\left(\frac{\tilde{\omega}}{\tilde{T}}\right)
\end{eqnarray}

\begin{eqnarray}
\label {eq.13}
\frac{N_0(\tilde{\omega})}{N}=1-\frac{N'(\tilde{\omega})}{N}
\end{eqnarray}

Fig.~5 illustrates temperature dependencies of the supracondensate particles  $N'$  and the particles in the condensate $N_0$  for the above-cited values of $\tilde{\omega}_c$ parameters.

\begin{figure}
\includegraphics{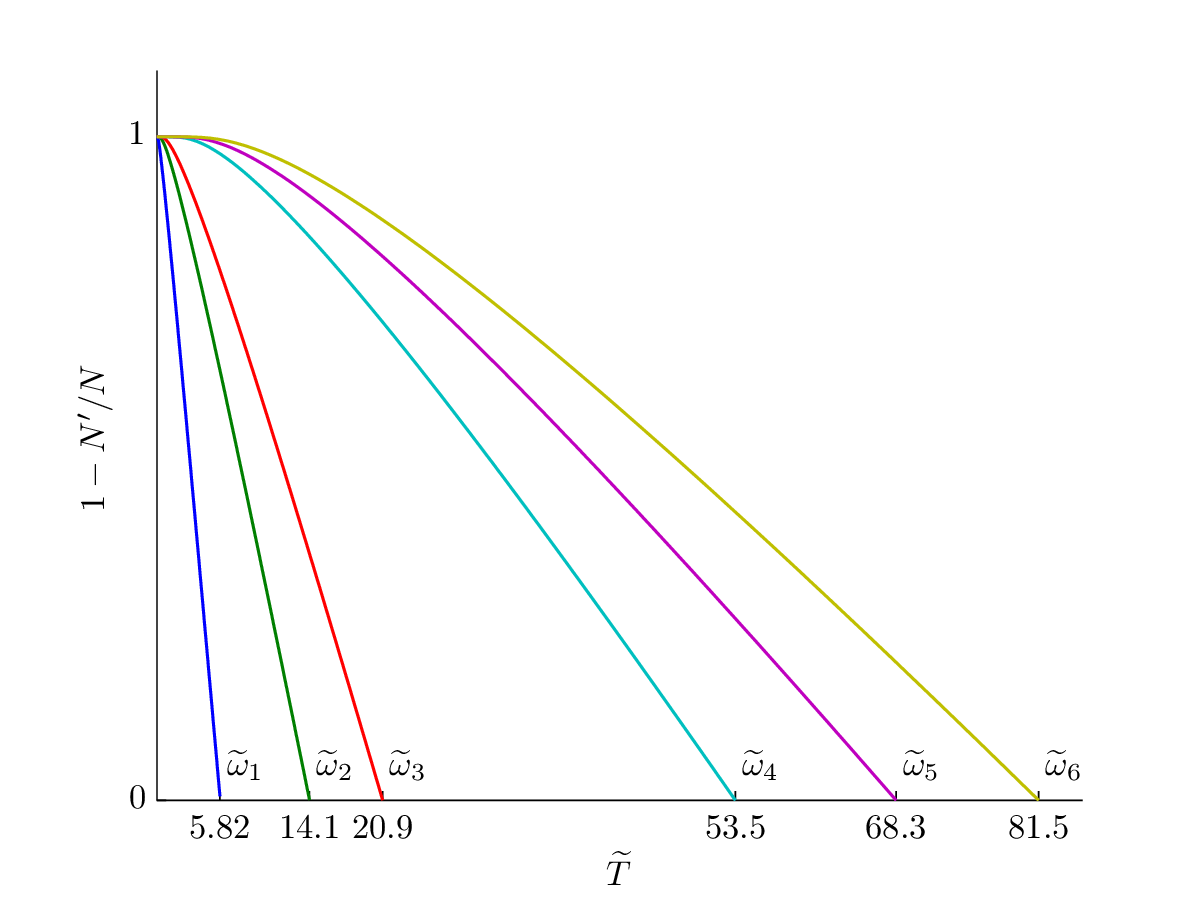}
\caption{Temperature dependencies of the relative number of supracondensate particles  $N'/N$   and condensate particles $N_0/N$
for the values of  $\tilde{\omega}_i$, given in Fig.~4.}
\end{figure}

From Fig.~5 it follows that, as we might expect, the number of particles in the condensate grows as the gap  $\omega _i$ increases.

The energy of TI-bipolaron gas $E$  reads:
\begin{eqnarray}
\label {eq.14}
E=\sum_{n=0,1,2...}\bar{m}_nE_n=E_{bp}N_0+\sum_{n\neq0}\bar{m}_nE_n
\end{eqnarray}

With the use of (37), (38) for the specific energy (i.e. energy per one TI- bipolaron) $\tilde{E}(\tilde{T})=E/N\omega^*$, $\tilde{E}_{bp}=E_{bp}/\omega^*$ (46) transforms into:
\begin{eqnarray}
\label {eq.15}
\tilde{E}=\tilde{E}_{bp}+\Delta\tilde{E}
\end{eqnarray}

\begin{eqnarray}
\label {eq.16}
\Delta\tilde{E}=\frac{\tilde{T}^{3/2}}{C_{1D}}F_{1/2}\left(\frac{\tilde{\omega}-\tilde{\mu}}{\tilde{T}}\right)
\left[\frac{\tilde{\omega}}{\tilde{T}}+\frac{F_{3/2}\left(\frac{\tilde{\omega}-\tilde{\mu}}{\tilde{T}}\right)}
{F_{1/2}\left(\frac{\tilde{\omega}-\tilde{\mu}}{\tilde{T}}\right)}\right],
\end{eqnarray}

\begin{eqnarray}
\label {eq.17}
F_{3/2}(\alpha)=\int^{\infty}_0\frac{\sqrt{x}dx}{e^{x+\alpha}-1}
\end{eqnarray}

where $\mu$   is determined by the equation:
\begin{eqnarray}
\label {eq.18}
C_{1D}=\tilde{T}^{1/2}_c F_{1/2}\left(\frac{\tilde{\omega}-\tilde{\mu}(\tilde{T})}{\tilde{T}}\right)
\end{eqnarray}

$$\tilde{\mu}=
\begin{cases}
0, & \tilde{T}<\tilde{T}_c; \\
\tilde{\mu}(\tilde{T}), & \tilde{T}>\tilde{T}_c.
\end{cases}
$$

Relation between  $\tilde{\mu}$  and the chemical potential of the system $\mu$  is given by the expression  $\tilde{\mu}=(\mu-E_{bp})/\omega^*$. Formulae (49), (50) also yield expressions for  $\Omega$ - potential:  $\Omega=-2E$  and entropy  $S=-\partial\Omega/\partial T$  ($F=-2E$, $S=-\partial F/\partial T$).

Fig.~6 demonstrates temperature dependencies of  $\Delta\tilde{E} =\tilde{E}-\tilde{E}_{bp}$  for the above-cited values of  $\tilde{\omega}_i$. Salient points of  $\Delta\tilde{E}_i(\tilde{T})$ curves correspond to the values of critical temperatures  $T_{c_i}$.

\begin{figure}
\includegraphics{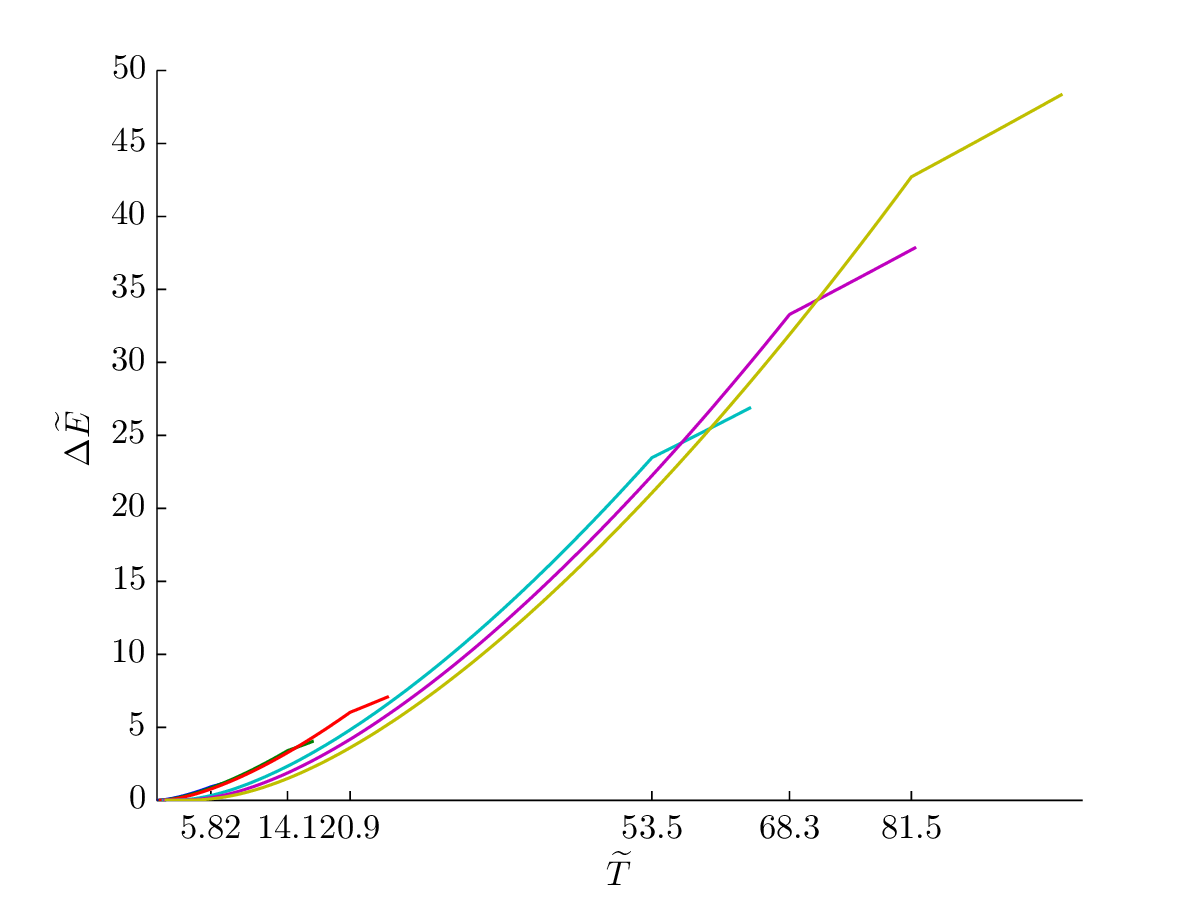}
\caption{The temperature dependencies $\Delta\tilde{E} =\tilde{E}(\tilde{T})-\tilde{E}_{bp}$ for various values of $\tilde{\omega}_i$ (see  Table~\ref{tab0}).}
\end{figure}

These dependencies enable us to find the heat capacity of TI-bipolaron gas:  $C_V(\tilde{T})=d\tilde{E}/d\tilde{T}$.

Fig.~7 shows temperature dependencies of the heat capacity $C_V(\tilde{T})$  for the above-cited values of  $\tilde{\omega}_i$. Table~\ref{tab0} lists the heat capacity jumps for the values of parameters $\tilde{\omega}_i$:
\begin{eqnarray}
\label {eq.19}
\Delta\frac{\partial C_V(\tilde{T})}{\partial\tilde{T}}=
{\frac{\partial C_V(\tilde{T})}{\partial\tilde{T}}}\Biggl|_{\tilde{T}=\tilde{T}_c+0}\Biggr.
-{\frac{\partial C_V(\tilde{T})}{\partial\tilde{T}}}\Biggl|_{\tilde{T}=\tilde{T}_c-0}\Biggr.
\end{eqnarray}

at the transition points.

\begin{figure}
\includegraphics{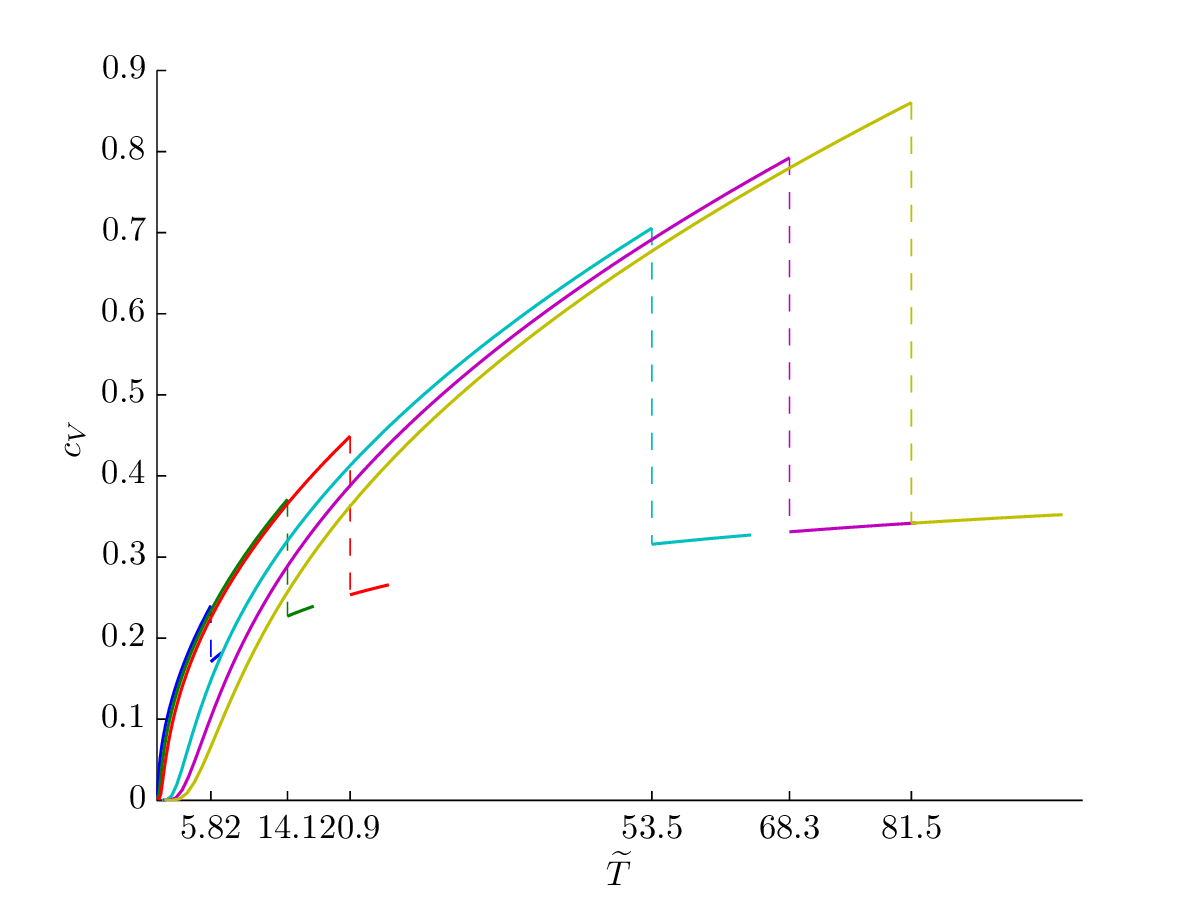}
\caption{The temperature dependencies of the heat capacity for various values of $\tilde{\omega}_i$ (see  Table~\ref{tab0}).}
\end{figure}

The dependencies obtained enable us to find the latent heat of the transition  $q=TS$, where $S$ is the entropy of supracondensate particles. At the transition point this value is  $q=2T_cC_V(T_c-0)$,$C_V=d\tilde{E}/d\tilde{T}$  and $\tilde{E}$   is determined by formulae (47), (48). The values of the heat of transition  $q_i$   for the above-cited values of $\tilde{\omega}_i$   are given in Table~\ref{tab0}.

\begin{table}
\caption{Dependence of critical temperatures  $\tilde{T}_{c_i}$, heat capacities  $C_V(\tilde{T}_{c_i}\pm0)$, and heat capacity jumps  $\Delta$ on the values of  $\tilde{\omega}_i$.}
\label{tab0}
\begin{tabular}{|c|c|c|c|c|c|c|}    \hline
\theadfont $i$ & 1& 2& 3& 4& 5& 6\\    \hline
$\tilde{\omega}_i$ & 0.2& 1& 2& 10& 15& 20\\  \hline
$\left.\tilde{T}_{c_i}\right.$ & 5.82& 14.11& 20.87& 53.47& 68.33& 81.5\\ \hline
$C_V(\tilde{T}_{c_i}-0)$ & 0.24& 0.37& 0.45& 0.71& 0.79& 0.86\\  \hline
$C_V(\tilde{T}_{c_i}+0)$ & 0.17& 0.23& 0.25& 0.32& 0.33& 0.34\\  \hline
$\frac{\partial C_V}{\partial\tilde{T}}(\tilde{T}_{c_i}-0)$ & $20.88\times10^{-3}$ & $13.51\times10^{-3}$ & $11.16\times10^{-3}$ & $7.11\times10^{-3}$ & $6.33\times10^{-3}$ & $5.83\times10^{-3}$ \\   \hline
$\frac{\partial C_V} {\partial\tilde{T}}(\tilde{T}_{c_i}+0) $ & $10.22\times10^{-3}$ & $4.72\times10^{-3}$ & $3.24\times10^{-3}$ & $1.19\times10^{-3}$& $0.89\times10^{-3}$& $0.73\times10^{-3}$\\    \hline
$\Delta$ & $-10.66\times10^{-3}$& $-8.79\times10^{-3}$ & $-7.93\times10^{-3}$ & $-5.92\times10^{-3}$ & $-5.44\times10^{-3}$ & $-5.11\times10^{-3}$ \\    \hline
\end{tabular}
\end{table}

\section {Comparison with discrete model}

	Earlier we considered the problem of symmetry breakdown for one electron interacting with oscillations of a one-dimensional quantum chain \cite{lit26}. According to Ref. \cite{lit26}, a rigorous  quantum-mechanical treatment leads to delocalized translation-invariant electron states, or to a lack of soliton-type solutions, breaking the initial symmetry of the Hamiltonian.
	
	In this paper we have shown that when the chain contains two electrons which interact with its oscillations and suffer Coulomb repulsion determined by the interaction, a stable state can be formed which does not violate translation invariance and has a lower energy  than the localized solution which breaks TI symmetry does.

Presently most papers describing electron states in discrete molecular chains are based
on Holstein-Hubbard Hamiltonian \cite{lit27}, \cite{lit33}, \cite{lit34}, \cite{lit-Korepin}:

\begin{eqnarray}\label{34}
H=\eta\sum_{i,j,\sigma ,\sigma '}c^+_{i\sigma}c_{j\sigma '}+\sum_i\hbar\omega\left(a^+_ja_j+1/2\right)+
\sum_jg\hat{n}_j\left(a^+_j+a_j\right)+\\\nonumber
\sum_{j,\sigma ,\sigma '}U\hat{n}_{j\sigma}\hat{n}_{j\sigma '},
\end{eqnarray}
where  $\hat{n}_{j\sigma}=c^+_{j\sigma}c_{j\sigma '}$, $\hat{n}_j=\sum_{\sigma}\hat{n}_{j\sigma}$,
$c^+_{j\sigma}$, $c_{j\sigma}$ - are operators of the birth and annihilation of an electron
with spin $\sigma$  at the  j-th site; $\eta$ - is the matrix element of the transition between nearest sites $(i, j)$.

Numerical investigations of Hamiltonian \eqref{34}  are based on the use of ansatz for the wave functions of the ground state:
\begin{eqnarray}\label{35}
\left|\Psi\right\rangle =\sum_{i,j,\sigma ,\sigma '}\Psi_{ij}c^+_{i\sigma}c^+_{j\sigma '}\left|0\right\rangle,
\end{eqnarray}
where  $\left|0\right\rangle$ - is a vacuum wave function which is a product of electron and lattice vacuum functions.

Hamiltonian (1) considered in this work is a continuum analog of Hamiltonian \eqref{34} if in \eqref{34} we put: $m=\hbar^2/2\eta a^2_0$,
 $\Gamma =Ua_0$. As is shown in Ref.~\cite{lit35}, presentation of the wave function as a product of the electron wave function by
the lattice one (Pekar ansatz) does not give an exact solution of Hamiltonian (1).
A similar conclusion is valid for Hamiltonian \eqref{34}.
In this context it would be interesting to discuss the limits of applicability of ansatz \eqref{35} in a discrete case using a particular example.

By way of example of a discrete model let us take the results of calculation of bipolaron states in
a Poly G/Poly C nucleotide chain given in paper \cite{lit36}. The Table~\ref{tab1} lists the values of the coupling energy
$\Delta=\left|E_{bp}-2E_p\right|$ in the case of  $U=0$, $\eta=0.084$ eV  for a discrete model \eqref{34}
with using ansatz \eqref{35}: $\Delta = \Delta^d$;  for a continuum Holstein
bipolaron with broken symmetry:  $\Delta = \Delta^H$ (6)-(7);
for a continuum TI-bipolaron: $\Delta = \Delta^{TI}$ (28).
These results suggest that $\Delta ^H$
 virtually coincides with $\Delta^d$ and becomes less than $\Delta^d$ as
$\kappa\leq0.1$ gets larger.
On the contrary, the values of $\Delta^{TI}$   for   $\kappa\leq0.3$   exceed the values of  $\Delta^d$
 and become less than $\Delta^d$ as  gets larger.
In the general case we can say that discreteness violates continual
translation invariance of the chain only when some threshold value of the coupling constant is exceeded.
In particular, for the discrete model of a Poly G/Poly C chain \cite{lit36} with parameters
$U \approx 1$eV, $\kappa=4g^2/\hbar\omega=0.5267$ which correspond, according to (8),
to $\gamma \approx 7.6$, the TI-bipolaron states considered in the paper are probably
unstable, since they do not fall on the stability interval $\gamma < \gamma _c$  (32).
In this case the states should be calculated based on a discrete model. For DNA,
such a calculation, as applied to the possibility of superconductivity in DNA
was carried out in papers \cite{lit16}, \cite{lit36}.
It should also be noted that apart from the condition $\gamma < \gamma _c$,
for continuum TI-bipolarons to exist, the condition of continuity should also be met.
According to Ref. \cite{lit35} it implies that the characteristic phonon vectors making the main
contribution into the energy of TI-bipolarons should satisfy the inequality $ka_0<1$.
From (27) it follows that the main contribution into the energy is given by the values of
$k \leq a$. For $U=0$, this yields $k \leq g^2/\hbar\omega\nu a_0$.
Accordingly, the condition of continuity takes on the form:

\begin{eqnarray}\label{36}
g^2/\hbar\omega\nu\leq 1
\end{eqnarray}
Obviously, for $U=0$, this condition is equivalent to the requirement $r/a_0\geq 1$,  where $r$ is determined by (6).
From (6) it also follows that for $U \neq0$ Holstein polaron becomes lengthier,
since its characteristic size becomes equal to $r = r_0(1-\gamma/4)^{-1}$,
where $r_0$  - is the characteristic size for $U=0$.
 For a TI-bipolaron, the same conclusion follows from expression (33) for the correlation length and Fig.~2.
Physically this is explained by the fact that Coulomb repulsion leads
to an increase of the characteristic distance between the electrons in the bipolaron state. Earlier this result was also obtained in Ref. \cite{lit30e}. Hence, though TI-bipolarons are delocalized, the requirements of continuity
for TI-bipolaron and Holstein bipolaron turn out to be similar.

\begin{table}
\caption{Coupling energies $\Delta$ for $U=0$ for a discrete model  $\Delta ^d$, for continuum Holstein model  $\Delta ^H$, and translation invariant bipolaron $\Delta ^{TI}$.}
\label{tab1}
\begin{tabular}{|c|c|c|c|c|c|}    \hline
\theadfont \diagbox [width=6em]{$\Delta$} {$\kappa$}&
 0.1 &  0.1975  &  0.296 &  0.359  &  0.5267 \\    \hline
	$\Delta ^d$ & 0.0037 & 0.015 & 0.05& 0.112 & 0.203 \\  \hline
	$\Delta ^H$ & 0.0037 & 0.0145 & 0.033  &   &  \\  \hline
	$\Delta ^{TI}$ & 0.0056 & 0.022 & 0.0495 &   &     \\  \hline	
\end{tabular}
\end{table}
The Table~\ref{tab1} lists the values of $\Delta$ for which the continuum model is more preferable than the 'exact' discrete one.

The results obtained suggest that  for parameter values when the continuum model is valid
and conditions of strong coupling are met, TI-bipolarons are energetically more advantageous.
Therewith the question of the character of a transition from the continuum description to the discrete one remains open.
One would expect that such a transition will occur with a sharp increase in the bipolaron effective mass as a result
of which the molecular chain will change from highly conducting state to low conducting one.

\section {Discussion of results}

The estimate of the value of the coupling constant $g_c=g/\hbar\omega_{0}$  sufficient for
the formation of translation-invariant bipolaron states in the region where the
criterion of their existence is met $0 < \gamma < \gamma_c$ can be obtained by comparing
the total energy of a strong coupling bipolaron with twice energy of individual
weak coupling polarons. Weak coupling polarons, by their treatment per se
(perturbation theory) are translation invariant with the energy \cite{lit28}:
\begin{eqnarray*}
E_p = -g^2\sqrt{ma^{2}_{0} / 2\hbar^3\omega_{0}}
\end{eqnarray*}

In particular, for $\gamma = 0$ we get: $g_c\approx 0.87(\hbar / ma^{2}_{0}\omega_{0})^{1/4}$. Hence, for the
overwhelming majority of various systems $g_c \leq 10$.

Notice that an application of an external magnetic field will cause the
decay of singlet bipolarons considered here since the energy of an individual
polaron in a magnetic field $H$ shifts by $-g_L \mu_B H/2$, where $g_L$ is Lande
factor, $\mu_B = |e|\hbar/2mc$ is a Bohr magneton. Being singlet, bipolarons do
not experience such a shift. Hence, the region of a bipolaron stability is
determined by the inequality $H < H_c$, where:

\begin{eqnarray*}
H_c = \frac{1}{\sqrt{2\pi}y_{min}(\gamma)}\left(\frac{\gamma_c - \gamma}{\gamma}\right) \frac{ma^{2}_{0}}{\hbar^2}\frac{g^4}{\hbar^2\omega^{2}_{0}}
\end{eqnarray*}
This estimate is valid for the case of non-quantizing magnetic fields.

As is known, the main mechanism leading to finite resistance in solid
bodies is dissipation of charge carriers on phonons \cite{lit-ziman}. In the case of translation
invariant bipolarons the separation of the system into bipolarons and optical
phonons is pointless. For a translation invariant bipolaron in the strong
coupling limit, the wave function of the system cannot be divided into electron
and phonon parts. The total momentum of a translation invariant bipolaron
is a conserving value, the relevant wave function is delocalized over the space
and a translation invariant bipolaron occurring in a system consisting only
of electrons and phonons, will be superconducting.
Inclusion of acoustical phonons into consideration leads to a limitation
on the possible value of the velocity $v$ of a translation invariant polaron or
bipolaron at which they have superconducting properties, namely, according
to the laws of energy and momentum conservation, this velocity should be
less than that of sound $s$. For $v > s$, a translation invariant polaron and
bipolaron become dissipative.

In a real system containing defects or structural imperfections with attractive
potential, these defects and imperfections will always trap polarons and
bipolarons with spontaneously broken symmetry. On the contrary translation
invariant bipolarons will form a bound state only if the potential well is
deep enough. Otherwise, even in an imperfect system, translation invariant
bipolarons will be delocalized. Notwithstanding the lack of bound states
in the presence of defects, the total momentum of a bipolaron no longer
commutates with the Hamiltonian and therefore is not an integral of the
system's motion. In this case a bipolaron will scatter elastically on a defect as
a result of which only its momentum will change. This scattering does not lead
to an energy loss. In the absence of dissipation the motion of bipolarons will
occur without friction and superconductivity in the system will be retained.
In the presence of large defects or imperfections possessing a great trapping
(scattering) potential, the system under discussion cannot be considered as
infinite any longer.

\section* {Conclusions}
In this paper we demonstrate that TI-bipolaron mechanism of Bose condensation can support superconductivity even for infinite chain. According to Fig.~6 the condensation in 1D systems is the phase transition of second kind.

The theory resolves the problem of the great value of the bipolaron effective mass. As a consequence, formal limitations on the value of the critical temperature of the transition are eliminated too. The theory quantitatively explains such thermodynamic properties of HTSC-conductors as availability  and value of the jump in the heat capacity lacking in the theory of Bose condensation of an ideal gas.
The theory also gives an insight into the occurrence of a great ratio between the width of the pseudogap and $T_c$. It accounts for the small value of the correlation length and explains the availability of a gap and a pseudogap in HTSC materials.

Accordingly, isotopic effect automatically follows from expression \eqref{eq.11}, where the phonon frequency $\omega_0$ acts as a gap.

Earlier the 3D TI-bipolaron theory was developed by author in \cite{Lak-2010}, \cite{Lak-2012}, \cite{Lak-2013}, \cite{lit35}. Consideration of 1D case carried out in the paper can be used to explain 3D high-temperature superconductors (3D TI-bipolaron theory of superconductivity was developed in Ref. \cite{litLak-2015}) where 1D stripes play a great role. As the consideration suggests, artificially created nanostripes with enhanced concentration of charge carriers can be used to increase the critical temperature of superconductors. Theoretical description of the nanostripes can also be based on the approach developed.

\section* {Declarations}
\subsection* {Acknowledgements}
The work was supported by projects RFBR N 16-07-00305 and RSF N 16-11-10163.

\end{document}